\documentclass[prb,preprint]{revtex4}
\usepackage{amsmath}  
\usepackage{amsfonts} 
\usepackage{graphicx} 
 \providecommand{\U}[1]{\protect\rule{.1in}{.1in}}
\begin{document}
\title{The Fluidyne engine}
\author{Alejandro Romanelli}
\altaffiliation{Electronic mail: alejo@fing.edu.uy}
\affiliation{Instituto de F\'{\i}sica, Facultad de Ingenier\'{\i}a\\
Universidad de la Rep\'ublica\\ C.C. 30, C.P. 11000, Montevideo, Uruguay}
\date{\today}
\begin{abstract}
The Fluidyne is a two-part hot-air engine, which has the peculiarity that both its power piston and displacer
are liquids. Both parts operate in tandem with the common working gas (air) transferring energy from the
displacer to the piston side, from which work is extracted. We describe analytically the thermodynamics of
the Fluidyne engine using the approach previously developed for the Stirling engine. We obtain explicit
expressions for the amplitude of the power piston movement and for the working gas temperatures and pressure
as functions of the engine parameters. We also study numerically the power and efficiency of the engine in
terms of the phase shift between the motions of piston and displacer.
\end{abstract}
\maketitle
\section{Introduction}
The Fluidyne engine was invented by Colin West in $1969$.\cite{Colin} It is a Stirling machine with one or more liquid pistons. Usually it contains air as
a working gas, and either two liquid pistons or one liquid piston and a displacer. The Fluidyne engine operates at a low frequency, typically $\sim1$ Hz,
and close to atmospheric pressure. The most common application of the liquid-piston system is in irrigation pumping, particularly for irrigation or
drainage pumping in places where electric power may not be available.\cite{Grupta} Nowadays, many commercial setups are available for specific applications
in agriculture, building services, drinking water and sanitation.\cite{Tom}

The basic principle of a Fluidyne engine is the fact that air expands when heated and contracts when cooled. It is possible to find many interesting videos
that present the construction and operation of this type of engine for demonstration and teaching purposes, on the internet. However it is not easy to find
theoretical material that explains its thermodynamics without an overly complicated technical discussion of the Stirling cycle. In the present paper we
study the thermodynamics of a particular type of Fluidyne engine that can be seen as a simple Stirling engine with one free liquid piston and
displacer.\cite{Walker,Darlington,Grupta}

Recently, we developed an alternative theoretical model\cite{alternativo} for the usual Stirling cycle.\cite{Zeman} The main characteristic of that
approach is the introduction of a polytropic process,\cite{romanelli} for which $PV^{\beta}= constant$, as a way to represent the exchange of heat with the
environment. The assumption of a polytropic process allows to model different dependencies of the working gas temperature with its volume. This means that
the polytropic index $\beta$ can be used as an additional degree of freedom that could be adjusted to the experimental data of a real operating engine. In
particular, we remark that all the discussion presented remains unchanged if $\beta$ is replaced by $\gamma={c_P}/{c_V}$ because this last is a special
case of the model.

Our alternative model provides analytical expressions for the pressure, temperatures  of the working gas and the work and heat exchanged with the heat
reservoirs. The theoretical pressure-volume diagram achieved a closer agreement with the experimental one than the standard analysis. Due to the generality
of the analytical expressions obtained, they can be adapted to any type of Stirling engine. In the present paper we use the mentioned model to study the
thermodynamics of the liquid-piston Stirling engine, ``the Fluidyne engine''.

The paper is organized as follows. In the next section we present the liquid-piston Stirling engine and we study in detail the dynamics of the liquid
piston with the help of the results of the thermodynamic model. In the third section we obtain the pressure-volume and the work-efficiency diagrams for the
Fluidyne engine. In the last section we present the main conclusions.

\section{Liquid piston Stirling engine}
\begin{figure}[h]
\begin{center}
\includegraphics[scale=0.65]{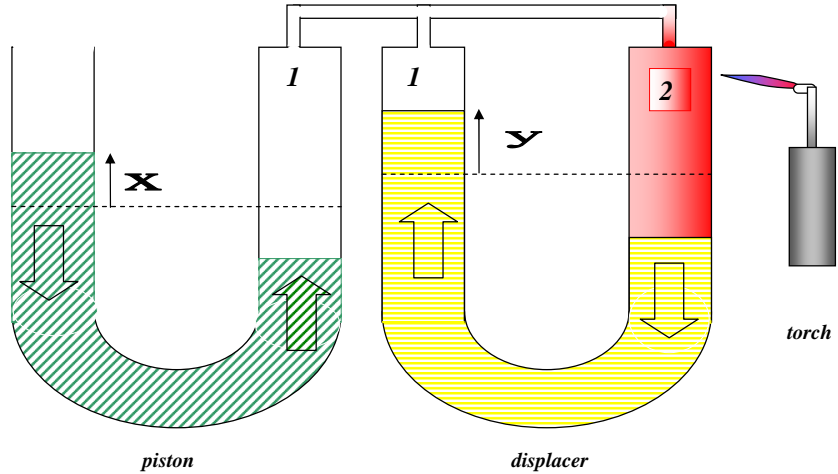}
\end{center}
\caption{Simple model of a Fluidyne engine. The piston and displacer are liquids, the cylinders are two U-tubes. The working gas moves in zones $1$ and $2$
that are, respectively, the cold and hot zones. Both zones have the same pressure but the open branch of the left U-tube is subjected to atmospheric
pressure. $x$ and $y$ indicate the displacement of the piston and the displacer from the their equilibrium positions. Open arrows indicate the movements of
piston and displacer. The torch works as a heat reservoir.} \label{f1}
\end{figure}

The Fluidyne engine consists of two U-tubes partially filled with liquid and connected with a tube of negligible caliber as shown in Fig.~\ref{f1}. One end
of the tube containing the piston liquid is open to the atmosphere. The three connected sections also contain the working gas. Figure~\ref{f1} shows a
snapshot picture of the engine working with its two liquids in motion modifying the gas volume. In this figure the liquids in the left and right U-tubes
are respectively the power piston and the displacer of the engine. While the gas volume changes due to the motion of the liquids it maintains a uniform
pressure determined by the action of the heat reservoir, represented by the torch in the figure. Two zones can be distinguished by their temperature in the
volume occupied by the gas and labeled with the numbers $1$ and $2$ in Fig.~\ref{f1}. Zone $2$ is the hot zone where the gas is in contact with the
reservoir at the external temperature $T_h$, and zone $1$ is the cool zone where the gas is in contact with the reservoir at the external temperature
$T_c$. Then, the working gas is never completely in either the hot or cold zone of the engine. However, the upper connection between the zones imposes the
same pressure $P$ in both. (Note that the diameter of the upper connection is relatively small but still very large compared with the mean free path of the
gas molecules, this condition ensures that there is no effusion process involved.)\cite{reif} Consequently, the gas density must be different in each zone
to keep the same pressure with different temperatures. The left side of the liquid piston is the free zone, here the pressure is the atmospheric pressure
$P_0$ and the temperature is also $T_c$.

In order to start the engine, we must set in motion externally both the piston and the displacer. One way to produce this initial motion starting from the
static situation, where both liquids are at rest, is that an external agent performs the following three steps: first it rotates the engine a small angle
with respect to an axis perpendicular to the main plane of the engine (the plane of Fig.~\ref{f1}), second it reverses the rotation to return the tubes to
the initial position and third it temporally manipulates the pressure over the left side of the liquid piston in such a way as to obtain the desired
relative movement between displacer and piston and simultaneously going back to the pressure $P_0$. After these steps the Fluidyne engine is in the
situation shown by Fig.~\ref{f1}, where both liquids are moving inside their U-tubes, and the only external agents that interact with the gas are the heat
reservoirs.

With the liquids in motion, the mechanism of energy transfer between the hot and cold zones works without any type of one-way valves and can be
qualitatively understood as follows. Initially we focus on the movement of the displacer and suppose the piston still. When the displacer is at the center
of the U-tube, suppose that the air volumes are more or less the same in the hot and cold zones. When the displacer liquid moves towards the cold (hot)
zone air is displaced, through the upper connecting tube, towards the hot (cold) zone and a greater quantity of air increases (decreases) its temperature
and consequently air pressure increases (decreases). Then the air pressure of the right side of the liquid piston (see Fig.~\ref{f1}) depends on the
displacer's motion but the left side remains at constant atmospheric pressure. If we now incorporate the motion of the piston, the net motion of air
between zones 1 and 2 depends on the phase of the relative motions of the displacer and piston fluids, which results in a cyclical transfer of energy from
the high temperature zone 2 to the piston liquid of zone 1 (with a small amount going into sustaining the oscillation of the displacer liquid against
dissipative losses). The phase of this motion is a crucial parameter for the efficiency and power delivered by the engine, and is treated in the analysis
that follows.

When the displacer liquid is set into oscillation in its U-tube, then the gas above this liquid is transferred back and forth between the hot and cold
zones. Both sides of the displacer tube have the same pressure, even when the engine is working, therefore if the displacer behaves as an ideal fluid any
initial oscillation never decreases. In what follows we assume that the displacer behaves as an ideal fluid subject to the restoring force of gravity, and
any oscillation has a natural frequency given by
\begin{equation}
\omega_d=\sqrt{\frac{2g}{L_d}},  \label{omega0}
\end{equation}
where $g$ is the gravity acceleration and $L_d$ is the displacer's length. Then the amplitude of the oscillation is given by
\begin{equation}
y=y_0 \sin \omega_d t, \label{newton1}
\end{equation}
where $y_0$ is the maximum amplitude and we take a vanishing initial phase. If the displacer is not an ideal fluid it is always possible to use part of the
energy produced by the engine to compensate losses and to maintain the original oscillation. As we shall see $\omega_d$ will be the operating frequency of
the engine.

To study the dynamics of the power piston we observe that the pressures on the right side $P$ and left side $P_0$ give rise to a net force on the liquid
piston proportional to $P-P_0$. We also consider an external force that dissipates all the useful power delivered by the working gas. In a real engine the
power piston should be coupled to a crankshaft to transfer the useful power. We opt for this simplified model in order to obtain a closed analytical
solution, however the real system can be addressed numerically. Therefore, the power piston dynamics is determined by Newton's second law applied to the
displacement $x$ of the liquid piston (see Fig.~\ref{f1})
\begin{equation}
\rho aL_p\ddot{x}=-b\dot{x}-{2g\rho a}x+(P-P_0)a, \label{newton}
\end{equation}
where $\rho$, $a$ and $L_p$ are the piston liquid density, cross-sectional area and length of fluid respectively and $b$ is a damping coefficient.

To obtain the working gas pressure $P$ we assume the following simplifying hypotheses for the working gas: (i) The gas behaves as a classical ideal gas.
(ii) The gas has uniform temperature in each zone namely $T_1$ and $T_2$. (iii) The mass of gas inside the thin connecting tubes is negligible, \emph{i.e.}
all the gas is inside the U-tubes where the displacer and the piston move. (iv) The expansion and compression processes are treated as polytropic
processes. For an ideal gas with a constant number of moles, the polytropic process may be defined through the relation $PV^{\beta}= constant$, where
$\beta$ is the polytropic index with typical values such that $1<\beta\leq \gamma$ with $\gamma=1.4$ for dry air at room temperature. In such a context we
have obtained in Ref.\onlinecite{alternativo} analytical expressions for  the pressure and temperatures of the working gas as functions of the gas volume,
the external temperatures and the polytropic index $\beta$ which characterizes the heat absorption process. The gas pressure is given by
\begin{equation}
P=P_{i}\left(\frac{\alpha\,V_{10}+V_{20}}{\alpha\,V_{1}+V_{2}}\right)\left(\frac{V_{0}}{V}\right)^{\beta-1},
\label{pe}
\end{equation}
and the temperatures $T_1$ and $T_2$ are
\begin{equation}
T_{1}=T_{c}\left(\frac{V_{0}}{V}\right)^{\beta-1},  \label{t1}
\end{equation}
\begin{equation}
T_{2}=T_{h}\left(\frac{V_{0}}{V}\right)^{\beta-1},  \label{t2}
\end{equation}
where $\alpha$ is the ratio between the temperatures of the reservoirs
\begin{equation}
\alpha\equiv{T_{h}}/{T_{c}}, \label{alfa1}
\end{equation}
$V_{1}$ is the gas volume of the cool zone with temperature $T_1$, $V_{2}$ is the gas volume of the hot zone with temperature $T_2$, $V=V_1+V_2$ is the
total  gas volume, $V_{10}$, $V_{20}$ and $V_0$ are initial conditions of $V_{1}$, $V_{2}$ and $V$ respectively, and $P_i$ is the initial condition of $P$.
It is interesting to underline that Eqs.~(\ref{pe}, \ref{t1}, \ref{t2}) for the isothermal case $\beta=1$ coincide with the so-called Schmidt solution,
published in $1871$.\cite{Schmidt}  For the case $\beta=\gamma$ the same equations describe the classical adiabatic solution.\cite{Bercho,Formosa} Both
examples show the ductility of polytropic processes to describe the thermodynamics of this type of engines.

Assuming that both U-tubes have the same cross section area $a$, the volumes of the zones $1$ and $2$ can be expressed as (see Fig.~\ref{f1})
\begin{equation}
V_1=V_{10}+a(x-y),  \label{v1}
\end{equation}
\begin{equation}
V_2=V_{20}+ay,  \label{v2}
\end{equation}
where the displacements $x$ and $y$ are taken from the initial (equilibrium) position, \emph{i.e.} $x=y=0$ give the initial volumes in Eqs.~(\ref{v1},
\ref{v2}). The total volume is then
\begin{equation}
V=V_{0}+ax.  \label{vtotal2}
\end{equation}
The explicit dependence of $P$ on $x$ and $y$ is obtained from Eqs.~(\ref{pe}), (\ref{v1}), (\ref{v2}) and (\ref{vtotal2}), then
\begin{equation}
P=\frac{P_i}{1+\frac{\alpha a}{\alpha V_{10}+V_{20}}[
x+(\frac{1}{\alpha}-1)y]}\left(\frac{1}{1+\frac{a}{V_{0}}x}\right)^{\beta-1}.
\label{pe2}
\end{equation}
It is clear now that Eq.~(\ref{newton}) has a non-linear dependence with both $x$ and $y$. However the engine operates in a closed regenerative cycle,
which presupposes only one characteristic frequency. Then, only the periodic solutions of Eq.~(\ref{newton}), with $\omega_d$ as a characteristic
frequency, will be useful. Here $\alpha\geq1$ and we further assume that the engine geometry is such that both $\frac{ax}{V_{10}}\ll 1$ and
$\frac{ay}{V_{10}}\ll 1$ are satisfied, and then these magnitudes can be treated as perturbations in Eq.~(\ref{pe2}). Expanding equation  Eq.~(\ref{pe2})
around the unperturbed volumes and substituting in Eq.~(\ref{newton}) the following linear equation is obtained
\begin{equation}
\ddot{x}+\frac{\xi}{\omega_d}\dot{x}+{\omega_p}^{2}x=\frac{P_i-P_0}{\rho
L_p}+\frac{aP_i(\alpha-1)}{\rho L_p(\alpha V_{10}+V_{20})}y,
\label{newton2}
\end{equation}
where
\begin{equation}
\xi=\frac{b\omega_d}{\rho L_pa}, \label{q}
\end{equation}
\begin{equation}
{\omega_p}^2=\frac{2g}{L_p}+\frac{aP_i}{\rho L_p}\left(
\frac{\beta-1}{V_{0}}+\frac{\alpha}{\alpha V_{10}+V_{20}}\right),
\label{w0}
\end{equation}
and $y$ is given by Eq.~(\ref{newton1}).

Equation~(\ref{newton2}) describes a linearly damped oscillator with external forcing. The first term on the right-hand side produces a constant shift from
the equilibrium position, which is equivalent to a redefinition of the volume $V_{10}$. Therefore, from now on we assume $P_i=P_0$ in order to ignore it
and we will concentrate on the periodic solution.

When the engine achieves the steady state motion, the solution of Eq.~(\ref{newton2}) that determines its
dynamics is
\begin{equation}
x=x_0 \sin(\omega_d t-\phi), \label{newton3}
\end{equation}
where
\begin{equation}
x_0=\frac{aP_0 (\alpha-1)}{\rho L_p\xi(\alpha
V_{10}+V_{20})}y_0\sin\phi, \label{x0}
\end{equation}
and
\begin{equation}
\phi=\arctan\left(\frac{\xi}{{\omega_p}^2-\omega_d^2}\right).
\label{phi}
\end{equation}
From Eq.(\ref{x0}) it is clear that the engine does not work if $\alpha=1$, that is, in order to function the engine needs heat reservoirs with $T_h>T_c$,
see Eq.(\ref{alfa1}). Moreover, when $\alpha\rightarrow\infty$ then $x_0$ tends asymptotically to a finite value. On the other hand, if $\omega_p=\omega_d$
in Eq.(\ref{phi}) the power piston is in resonance with the displacer, $\phi=\pi/2$, and the amplitude $x_0$ is maximum.
\section{Fluidyne thermodynamics}
\begin{figure}[t]
\begin{center}
\includegraphics[scale=0.4]{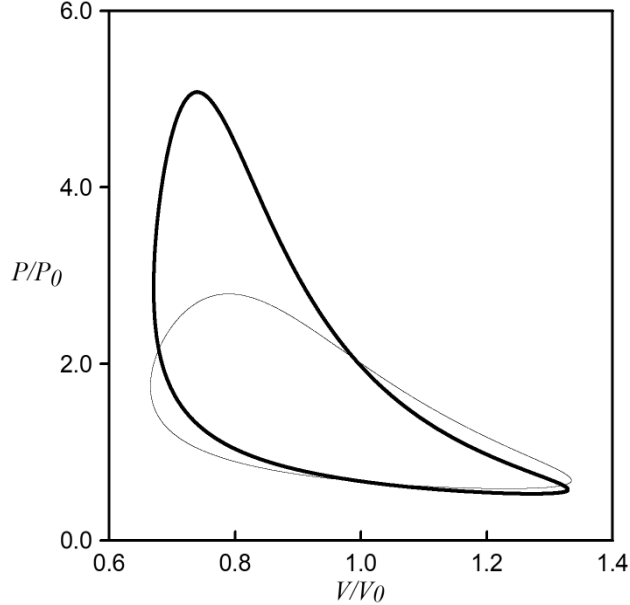}
\end{center}
\caption{Dimensionless pressure-volume diagrams for the gas of the Fluidyne engine; in thick line for $\phi=0.59 \pi$ and in thin line for $\phi=0.43 \pi$.
The parameters are $\alpha=10$, $\beta=1.25$, $V_{10}=1.0$, $V_{20}=0.5$, $y_0=0.60$ and $x_0=0.51\sin \phi$.} \label{f2}
\end{figure}

To obtain the pressure-volume diagram for the Fluidyne engine we use Eqs.~(\ref{newton1}), (\ref{pe2}) and (\ref{newton3}). This is shown in Fig.~\ref{f2}
for two characteristic values of $\phi$ that are to be explained below. The area inside such smooth closed curves represents the total work $W$ of the
cycle, whose expression is
\begin{equation}
W=\oint P\,dV=a\int_{0}^{\frac{2\pi}{\omega_d}}P\,\dot{x}\,dt,
\label{work}
\end{equation}
where the last equality follows from Eq.~(\ref{vtotal2}). Here $W$ is the work available for overcoming mechanical friction losses and for providing useful
power.

The differential equation for the heat absorbed by the working gas in the polytropic process was obtained in the general case of a Stirling engine in Ref.
 \onlinecite{alternativo} as follows:
\begin{equation}
dQ=\frac{P}{\gamma-1}\left[ (\gamma-\beta)dV+(\alpha -1)
\frac{V_1dV_2-V_2dV_1}{\alpha V_{1}+V_{2}}\right], \label{segunda}
\end{equation}
where $V_1$, $V_2$, $V$ are defined by Eqs.~(\ref{v1}), (\ref{v2}), (\ref{vtotal2}) and
\begin{equation}
\gamma=\frac{c_P}{c_V}, \label{gamma}
\end{equation}
is the quotient of the  specific heats of the gas at constant pressure $c_P$ and constant volume $c_{V}$. In this paper we take the polytropic index such
that $0\leq\beta\leq \gamma=1.4$.

It is important to emphasize that Eq.~(\ref{segunda}) together with Eqs.~(\ref{pe}), (\ref{t1}) and (\ref{t2}) determine the Fluidyne thermodynamics. These
four equations depend on the parameter $\alpha$ (the temperatures ratio) and they have finite asymptotic values when $\alpha\rightarrow\infty$. This means
that after a certain value of $\alpha$, no matter how much we increase the temperatures ratio, the pressure and the absorbed heat are bounded, and this in
turn explains why the useful work and the efficiency in any Stirling engine are asymptotically bounded.
\begin{figure}[h]
\begin{center}
\includegraphics[scale=0.4]{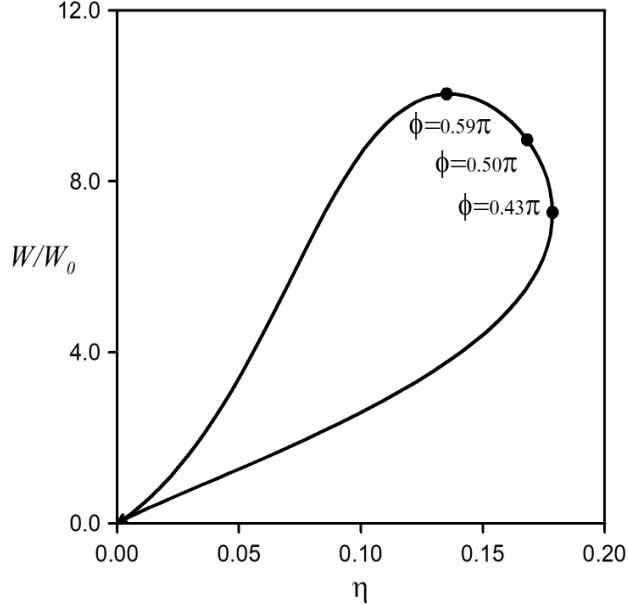}
\end{center}
\caption{Dimensionless work as a function of the efficiency where $W_0=P_0V_{10}$ and the angle $\phi$ varying between $0$ and $\pi$. The dots on the curve
indicate the values of $\phi$ for maximum efficiency ($\phi=0.43\pi$), resonance ($\phi=0.50\pi$) and maximum ($\phi=0.59\pi$) work. The values of the
parameters are the same as in Figure ~\ref{f2}.} \label{f3}
\end{figure}
\begin{figure}[ht]
\begin{center}
\includegraphics[scale=0.4]{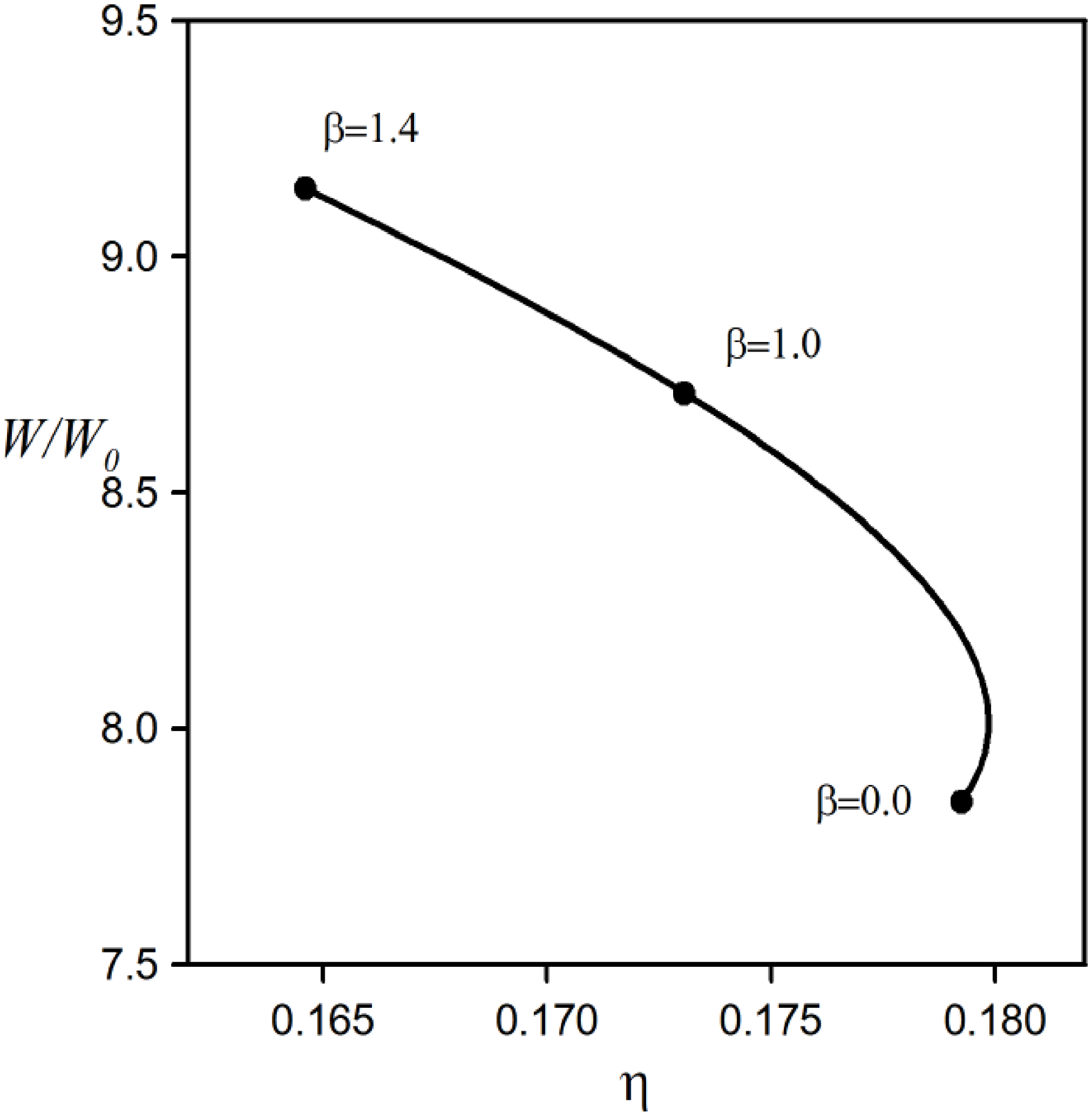}
\end{center}
\caption{Dimensionless work as a function of the efficiency when the polytropic index $\beta$ varies between $0.0$ and $1.4$. The dots on the curve
indicate three particular values of $\beta$. The other parameters are the same as those of previous figures. } \label{f9}
\end{figure}
Let us call $Q_{in}$ the heat absorbed by the gas in the cycle with the convention $Q_{in}>0$; similarly we call $Q_{out}<0$  the heat rejected. They can
be calculated numerically using Eq.~(\ref{segunda}) and the definitions
\begin{equation}
Q_{in}=\int_{\mathcal{C}_{in}}dQ, \label{c1}
\end{equation}
\begin{equation}
Q_{out}=\int_{\mathcal{C}_{out}}dQ, \label{c2}
\end{equation}
where $\mathcal{C}_{in}$ and $\mathcal{C}_{out}$ refer to the paths where $d{Q}>0$ and $d{Q}<0$ respectively. As the internal energy change in the entire
cycle vanishes, the total work verifies $W=Q_{in}+Q_{out}$. Therefore, the efficiency $\eta$ defined as $W/Q_{in}$ can be written as
\begin{equation}
\eta=1+{Q_{out}}/{Q_{in}}. \label{eta}
\end{equation}

We have integrated numerically Eqs.(\ref{work}), (\ref{c1}) and (\ref{c2}) using the standard Simpson's rule. Figure~\ref{f3} shows the relation between
work and efficiency as functions of the phase $\phi$. From this figure is clear that the phasing between the piston and the displacer plays a central role
in the functioning of the engine; in particular we observe that the maximum efficiency phase does not coincide with the maximum work phase. The $PV$
diagrams in Fig.~\ref{f2} correspond precisely to the phases of maximum work and maximum efficiency.

When $\phi=\pi/2$ the piston is in resonance with the displacer and this situation is intermediate between maximum work and maximum efficiency; there the
engine is in a good performance zone, where the engine operates in a good compromise between power and efficiency.

Figure~\ref{f9} shows the behavior of work and efficiency when the polytropic index $\beta$ varies. This parameter depends on the gas initial thermodynamic
conditions and in particular on its relative humidity.\cite{romanelli} The figure shows that the dependence  with  $\beta$ is stronger for the power than
for the efficiency.
\section{Conclusions}
Research on Stirling engines is one of the lines that contribute both to the rational use of energy and to sustainable development. In particular the solar
thermal conversion systems based on these engines are amongst the most interesting and promising research lines.
\cite{tesis1,tesis2,tesis3,tesis4,tesis5,tesis6} This paper presents an unusual application of heat engine thermodynamics combined with the physics of
oscillators at a level appropriate for advanced undergraduate students.

From a strictly technical point of view we have proposed a theoretical model that describes the thermodynamics of a Stirling engine in a simple, precise
and natural way.\cite{alternativo} The engine studied in this paper is a special type of the Stirling engine, therefore we have applied the theoretical
model to investigate its thermodynamics and dynamical evolution. The exchange of heat with the environment is modeled as a polytropic process. We obtain
analytical expressions for the piston amplitude, the working gas temperatures and pressure. These quantities are expressed as functions of (i) the phase
difference between the power piston and the displacer, $\phi$; (ii) the ratio of the temperatures of the heat reservoirs, $\alpha$; (iii) the polytropic
index, $\beta$ and (iv) the geometrical characteristics of the engine. We also study numerically the engine power and its efficiency  as a function of
$\phi$ and $\beta$ and show that, when in resonance, the engine works a good performance regime. These results show the versatility of the thermodynamic
model developed here to describe different types of Stirling engines and they encourage further work on this line.

In summary, the theoretical approach proposed describes analytically the Fluidyne engine thermodynamics and it contributes to an understanding of the
subtle way through which this engine transforms heat into work.

\section*{ACKNOWLEDGMENTS}
I acknowledge the stimulating discussions with V\'{\i}ctor Micenmacher and Raul Donangelo and the support from ANII and PEDECIBA (Uruguay).

\end{document}